\documentclass[sigconf,nonacm]{acmart}
\usepackage{graphicx} 

\title{Automatic Metadata Capture and Processing for High-Performance Workflows}
\author{Polina Shpilker}
\orcid{0000-0002-6761-7326}
\email{polina.shpilker@tufts.edu}
\affiliation{%
  \institution{Tufts University}
  \city{Medford}
  \state{Massachusetts}
  \country{USA}
}

\author{Line Pouchard}
\email{lcpouch@sandia.gov}
\orcid{0000-0002-2120-6521}
\affiliation{%
  \institution{Sandia National Laboratories}
  \city{Albuquerque}
  \state{NM}
  \country{USA}
}

\date{April 2025}

\begin{document}

\maketitle

\section{Abstract}

Modern workflows run on increasingly heterogeneous computing architectures and with this heterogeneity comes additional complexity. We aim to apply the FAIR principles for research reproducibility by developing software to collect metadata annotations for workflows run on HPC systems. We experiment with two possible formats to uniformly store these metadata, and reorganize the collected metadata to be as easy to use as possible for researchers studying their workflow performance.

\section{Motivation}

Scientific computing running on high performance computing systems (HPC) increasingly uses workflows to launch simulations and analysis, communicate with schedulers and automate resource allocation. These workflows strive to take advantage of the multitude of resources available to them on these powerful compute systems: tasks may be massively parallelized, memory-intensive scripts may run without restriction, and complex mathematical calculations can be offloaded onto GPUs.Under the latest generation of systems, the runtime environments of these workflows have become extremely heterogeneous. A given workflow can be run on many different types of hardware at once, different workflows require different levels of parallelization, and even the data accessed by the workflows themselves could be spread across different types of file systems. The heterogeneity increases the complexity of metadata annotations for a given workflow, much of the metadata that can be captured is outside a researcher's control, and the processes to capture and establish the provenance of workflow execution are themselves not unified or even missing.

\section{Approach}
\begin{figure}
    \centering
    \includegraphics[width=0.75\linewidth]{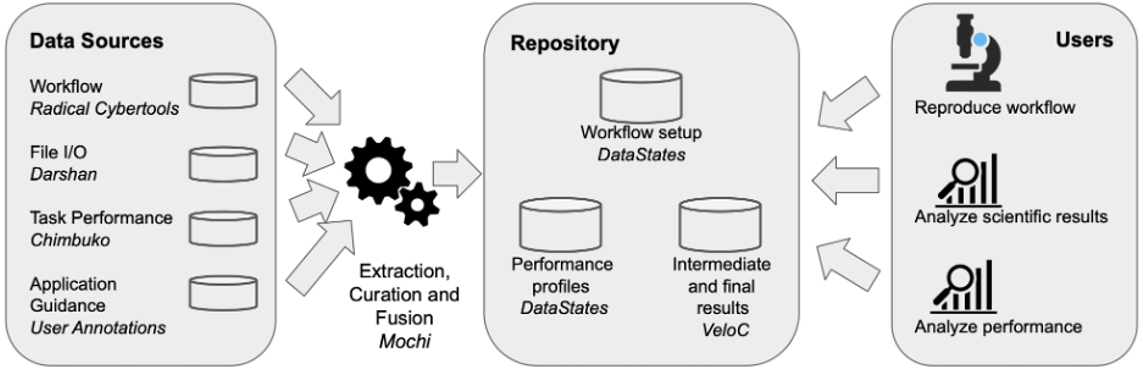}
    \caption{Graphical representation of the composable components that make up a RECUP workflow.}
    \label{fig:RECUP}
\end{figure}

We offer an approach from the software perspective to ensure that metadata is collected throughout the life cycle of a workflow running on a US DOE Leadership Computing Facility (LCF). To illustrate our approach we take the example of the RECUP\footnote{https://sites.google.com/view/recup-reproducibility/home\\ http://github.com/RECUP-DOE/} workflows, executed on the Polaris system hosted by Argonne National Laboratory (Fig. \ref{fig:RECUP})\cite{nicolae2023building}. In this example, workflows are defined and run using tools such as Radical Cyber Tools Pilot and Dask. File I/O data is generated by the I/O characterization tool Darshan\cite{PyDarshan}, and anomalous performance events are collected by the HPC anomaly detection system Chimbuko\cite{Chimbuko}. Outputs are collected by tools in the Mochi\cite{Mochi} portfolio for HPC data management and collected in a variety of repositories and analyzed by researchers.

We focus on the Data Sources portion of the pipeline to extract metadata related to performance reproducibility, defined in terms of minimal run-to-run variation in execution times of applications on the same system \cite{patki2019performance}. In order to assess workflow performance reproducibility, the workflow must be fully annotated by metadata. These sources each relate to a different component of the workflow, and each has its own unique file format, its own unique level of granularity, and its own software dependency. The lack of congruence between metadata from different sources poses a challenge that we address by experimenting with different methods for processing and storage, including RO-Crate for Chimbuko anomalous event metadata and Data Frames for Dask workflows and Darshan I/O.

\section{Extracting from Various Sources}

\subsection{Anomalous Event Metadata}

Anomalous event metadata related to performance anomalies are stored in the Chimbuko provenance database. These database entries completely describe the identity of the function that returned the anomalous event, including start and end time and the file containing the function definition. We investigated the Workflow Provenance Run RO-Crate Profile to package this information \cite{WPRRO-Crate} as it encapsulates FAIR principles.
 
The purpose of a Workflow Provenance Run RO-Crate is to describe the prospective provenance (the planned actions) directly beside and in relationship to the retrospective provenance (the actions as they occurred.) Using the information from the Chimbuko provenance database, we are able to span both of these categories to partially fill the information expected by this definition.

However, complete annotation of a Workflow Provenance Run RO-Crate proved impossible to fill using only metadata available from the Chimbuko provenance database, and would need to be added manually by users. This defeats the purpose of automated metadata capture and processing for our workflows as the Chimbuko provenance database doesn't contain such information. While RO-Crate is well-suited to FAIR workflows and workflow management systems that are capable of providing complete information, in our case, RO-Crate proved an inappropriate target for further metadata collection.

\subsection{Workflow Event Metadata}

Our second target of capture is workflow metadata. These metadata describe workflow behavior during execution, e.g the workflow execution patterns, and workflow performance metadata that reports on a workflow system overhead. RECUP supports DASK, a python library for running massively parallelized workflows on HPC systems. The DASK-Mofka coupler enables the collection of worker and scheduler transition events\cite{DaskMofka} and generates three files: one for scheduler-level transitions, one for worker-level transitions, and one for worker file transfers. 

These files are event-focused, but latest generation HPC workflows are task-focused: the point of reference in a workflow is the task or line of python code that caused an event. To shift from this event-focused paradigm to a task-focused paradigm, we developed software that processes these metadata collections into Task objects. These objects can be more easily searched by researchers to identify events related to a particular step in their workflow, thus pointing to a possible cause of performance variation.
  
\subsection{I/O Metadata}

In the execution of a workflow, Darshan creates I/O log files for every spawned process. Each of these log files contains several modules, such as POSIX and LUSTRE modules. In order to make these metadata more accessible to researchers, we used the PyDarshan python library to read these logs and translate them into pandas dataframes.

These logs are still difficult for researchers to use due to being spread out over hundreds of individual dataframes. We reorganize these dataframes to group them by module across processes, as well as combine dataframes that come from the same module (such as POSIX read and write segment logs.) The resulting 5-6 (one per module logged by Darshan) dataframes are then written to disk in a way that researchers can easily re-import them as dataframes. The re-organization into fewer files enables easy and granular searching of execution events, supporting FAIR in performance studies.

\section{Metadata Alignment and Future Work}

Collected metadata from the above sources need to be exported into a singular representation to ease loading into analysis tools. Currently, Chimbuko anomalous event metadata is collected in an RO-Crate format based on JSON-LD. The Dask-Mofka workflow event metadata is stored in custom classes which are then pickled and saved onto disc. The Darshan I/O metadata is contained in python pandas dataframes and then saved to disk using Apache Parquet.  This is the closest we have to a uniform representation.  

The Dask-Mofka workflow metadata can easily be converted into dataframes and stored beside the Darshan I/O logs. The Chimbuko anomalous event metadata requires additional work to align it with the Dask-Mofka metadata, as the anomalous events are most informative in the context of a workflow task. While our work greatly facilitates the precise and cross-source identification of performance events and their provenance, much work remains to be done to make these workflows FAIR. This work demonstrates the non-trivial processing required to automatically extract and reconcile metadata from composable workflow tools in HPC, and lays a metadata foundation for end-to-end performance analysis studies of workflow executions.




\section*{Acknowledgments}
This manuscript has been authored in part by employees of Sandia National Laboratories operated by NTESS for the U.S. DOE/NNSA under contract DE-NA0003525.

\bibliography{refs}
\bibliographystyle{plainnat}
\end{document}